\documentclass[twoside,twocolumn,9pt]{article}
\usepackage{extsizes}
\usepackage[super,sort&compress,comma]{natbib} 
\usepackage[version=3]{mhchem}
\usepackage[left=1.5cm, right=1.5cm, top=1.785cm, bottom=2.0cm]{geometry}
\usepackage{balance}
\usepackage{times,mathptmx}
\usepackage{sectsty}
\usepackage{graphicx} 
\usepackage{lastpage}
\usepackage[format=plain,justification=justified,singlelinecheck=false,font={stretch=1.125,small,sf},labelfont=bf,labelsep=space]{caption}
\usepackage{float}
\usepackage{fancyhdr}
\usepackage{fnpos}
\usepackage[english]{babel}
\addto{\captionsenglish}{%
  
}
\usepackage{array}
\usepackage{droidsans}
\usepackage{charter}
\usepackage[T1]{fontenc}
\usepackage[usenames,dvipsnames]{xcolor}
\usepackage{setspace}
\usepackage[compact]{titlesec}
\usepackage{hyperref}

\usepackage[utf8]{inputenc} 
\usepackage{stix} 
\usepackage[dvipsnames]{xcolor}
\usepackage{soul} 

\newcommand{\celsius}{\,$^{\circ}$C\,}

\usepackage{epstopdf}

\definecolor{cream}{RGB}{222,217,201}

\begin{document}

\pagestyle{fancy}
\thispagestyle{plain}
\fancypagestyle{plain}{

\fancyhead[C]{\includegraphics[width=18.5cm]{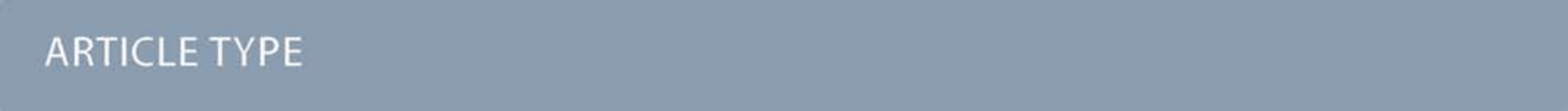}}
\fancyhead[L]{\hspace{0cm}\vspace{1.5cm}\includegraphics[height=30pt]{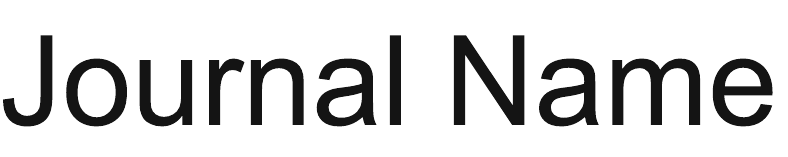}}
\fancyhead[R]{\hspace{0cm}\vspace{1.7cm}\includegraphics[height=55pt]{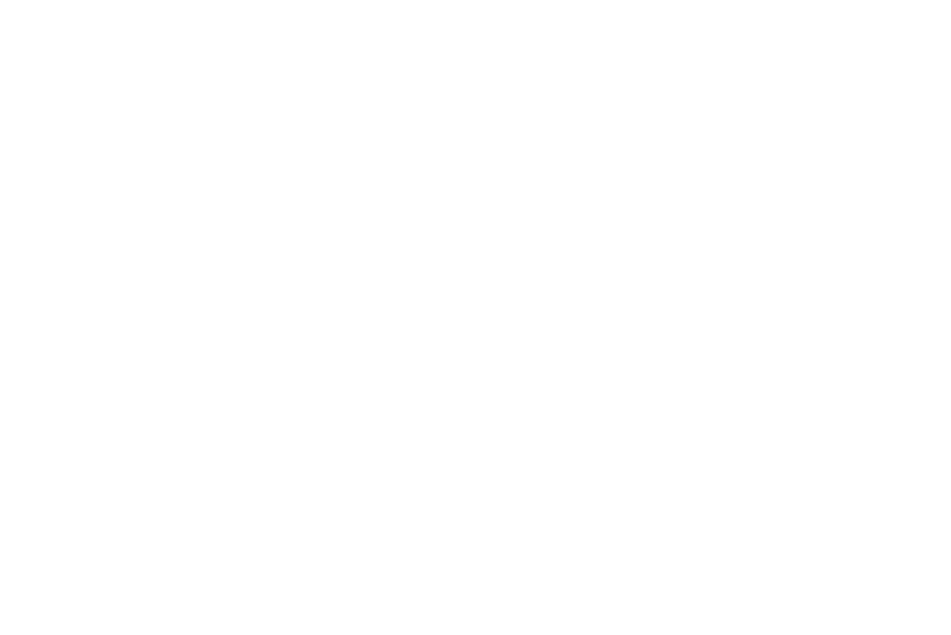}}
\renewcommand{\headrulewidth}{0pt}
}

\makeFNbottom
\makeatletter
\renewcommand\LARGE{\@setfontsize\LARGE{15pt}{17}}
\renewcommand\Large{\@setfontsize\Large{12pt}{14}}
\renewcommand\large{\@setfontsize\large{10pt}{12}}
\renewcommand\footnotesize{\@setfontsize\footnotesize{7pt}{10}}
\makeatother

\renewcommand{\thefootnote}{\fnsymbol{footnote}}
\renewcommand\footnoterule{\vspace*{1pt}%
\color{cream}\hrule width 3.5in height 0.4pt \color{black}\vspace*{5pt}} 
\setcounter{secnumdepth}{5}

\makeatletter 
\renewcommand\@biblabel[1]{#1}            
\renewcommand\@makefntext[1]%
{\noindent\makebox[0pt][r]{\@thefnmark\,}#1}
\makeatother 
\renewcommand{\figurename}{\small{Fig.}~}
\sectionfont{\sffamily\Large}
\subsectionfont{\normalsize}
\subsubsectionfont{\bf}
\setstretch{1.125} 
\setlength{\skip\footins}{0.8cm}
\setlength{\footnotesep}{0.25cm}
\setlength{\jot}{10pt}
\titlespacing*{\section}{0pt}{4pt}{4pt}
\titlespacing*{\subsection}{0pt}{15pt}{1pt}

\fancyfoot{}
\fancyfoot[LO,RE]{\vspace{-7.1pt}}
\fancyfoot[CO]{\vspace{-7.1pt}\hspace{13.2cm}}
\fancyfoot[CE]{\vspace{-7.2pt}\hspace{-14.2cm}}
\fancyfoot[RO]{\footnotesize{\sffamily{1--\pageref{LastPage} ~\textbar  \hspace{2pt}\thepage}}}
\fancyfoot[LE]{\footnotesize{\sffamily{\thepage~\textbar\hspace{3.45cm} 1--\pageref{LastPage}}}}
\fancyhead{}
\renewcommand{\headrulewidth}{0pt} 
\renewcommand{\footrulewidth}{0pt}
\setlength{\arrayrulewidth}{1pt}
\setlength{\columnsep}{6.5mm}
\setlength\bibsep{1pt}

\makeatletter 
\newlength{\figrulesep} 
\setlength{\figrulesep}{0.5\textfloatsep} 

\newcommand{\topfigrule}{\vspace*{-1pt}%
\noindent{\color{cream}\rule[-\figrulesep]{\columnwidth}{1.5pt}} }

\newcommand{\botfigrule}{\vspace*{-2pt}%
\noindent{\color{cream}\rule[\figrulesep]{\columnwidth}{1.5pt}} }

\newcommand{\dblfigrule}{\vspace*{-1pt}%
\noindent{\color{cream}\rule[-\figrulesep]{\textwidth}{1.5pt}} }

\makeatother

\twocolumn[
  \begin{@twocolumnfalse}
\vspace{3cm}
\sffamily
\begin{tabular}{m{4.5cm} p{13.5cm} }

\noindent\footnotesize{Cite this: DOI: \href{https://doi.org/10.1039/C8NA00369F}{10.1039/C8NA00369F}} & \noindent\LARGE{\textbf{Top-down fabrication of ordered arrays of GaN nanowires by selective area sublimation}} \\
\vspace{0.3cm} & \vspace{0.3cm} \\

 & \noindent\large{Sergio Fernández-Garrido,$^{\ast}$\textit{$^{ab}$} Thomas Auzelle,\textit{$^{a}$} Jonas Lähnemann,\textit{$^{a}$} Kilian Wimmer,\textit{$^{a}$} Abbes Tahraoui,\textit{$^{a}$} and Oliver Brandt\textit{$^{a}$}} \\

 & \noindent\normalsize{We demonstrate the top-down fabrication of ordered arrays of GaN nanowires by selective area sublimation of pre-patterned GaN(0001) layers grown by hydride vapor phase epitaxy on Al$_{2}$O$_{3}$. Arrays with nanowire diameters and spacings ranging from 50 to 90~nm and 0.1 to 0.7~µm, respectively, are simultaneously produced under identical conditions. The sublimation process, carried out under high vacuum conditions, is analyzed \emph{in situ} by reflection high-energy electron diffraction and line-of-sight quadrupole mass spectromety. During the sublimation process, the GaN(0001) surface vanishes, giving way to the formation of semi-polar $\lbrace1\bar{1}03\rbrace$ facets which decompose congruently following an Arrhenius temperature dependence with an activation energy of ($3.54 \pm 0.07$)~eV and an exponential prefactor of $1.58\times10^{31}$~atoms\,cm$^{-2}$\,s$^{-1}$. The analysis of the samples by low-temperature cathodoluminescence spectroscopy reveals that, in contrast to dry etching, the sublimation process does not introduce nonradiative recombination centers at the nanowire sidewalls. This technique is suitable for the top-down fabrication of a variety of ordered nanostructures, and could possibly be extended to other material systems with similar crystallographic properties such as ZnO.} \\

\end{tabular}

 \end{@twocolumnfalse} \vspace{0.6cm}

  ]

\renewcommand*\rmdefault{bch}\normalfont\upshape
\rmfamily
\section*{}
\vspace{-1cm}


\footnotetext{\textit{$^{a}$~Paul-Drude-Institut für Festkörperelektronik, Leibniz-Institut im Forschungsverbund Berlin e.V., Hausvogteiplatz 5–7, 10117 Berlin, Germany; E-mail: sergio.fernandezg@uam.es}}
\footnotetext{\textit{$^{b}$~Grupo de Electrónica y Semiconductores, Dpto.\ Física Aplicada, Universidad Autónoma de Madrid, C/ Francisco Tomás y Valiente 7, 28049 Madrid, Spain}}

\footnotetext{\dag~Electronic Supplementary Information (ESI) available: Analysis of GaN facets formed during thermal sublimation by reflection high-energy electron diffraction. See DOI: \href{https://doi.org/10.1039/C8NA00369F}{10.1039/C8NA00369F}}



\section{Introduction}
Since the first reports on the self-assembled formation of single-crystalline GaN nanowires (NWs) in the late nineties,\cite{Yoshizawa_jjoap_1997,Sanchez-Garcia_jcg_1998} increasing efforts have been devoted to master the bottom-up fabrication of random and ordered arrays of group-III nitride NW heterostructures by metal-organic chemical vapor deposition and molecular beam epitaxy.\cite{Li_JAP_2012} The research in this field is motivated by the technological relevance of GaN, the material that has enabled the commercialization of solid-state lighting for general illumination, and the potential advantages promised by the NW architecture, specifically: (i) the absence of threading dislocations (TDs) in heteroepitaxial growth, (ii) the possibility of combining lattice mismatched compounds without introducing extended defects, (iii) the absence (reduction) of the quantum-confined Stark effect in radial (axial) NW heterostructures, (iv) the improved light extraction efficiency in comparison to as-grown planar devices, and (v) the opportunity to increase the size of the active region per area unit of the substrate when fabricating devices with a core-shell geometry. \cite{Zubia_JAP_1999,Hersee_JMR_2011,Li_JAP_2012,Yue_IEEE_JQE_2013,Reddy_Nanotech_2016_01,Hauswald_ACS_Phot_2017_01} However, despite all these advantages and worldwide research activities, the performance of NW based light-emitting diodes (LEDs) lags behind that of their state-of-the-art two-dimensional counterparts fabricated on Al$_{2}$O$_{3}$ and Si substrates, which nowadays exhibit wall-plug and external quantum efficiencies above 80~\% in the blue spectral range.\cite{Nakagawa_JJAP_2013_01,Kimura_JAP_2016_01} The inferior performance of NW-based LEDs is mainly caused by the limitations and complexities inherent to their formation using bottom-up methods. Particularly, the nonideal growth conditions often required to promote either uniaxial or radial growth may favor the incorporation of higher concentrations of impurities and defects.\cite{Li_JAP_2012,Mohajerani_JJAP_2016_01} Furthermore, the different chemical and physical properties of the precursors, which are simultaneously deposited on different crystal facets, can result in the formation of detrimental compositional and structural inhomogenities.\cite{Albert_Cryst_Growth_Des_2015,Mohajerani_JJAP_2016_01,Muller_Nano_Lett_2016_01}

Given that the epitaxial growth of group-III nitrides has reached a high level of maturity, it is appealing to combine this well-developed technology with some of the advantages offered by the NW architecture using a top-down approach. \cite{Chiu_Nanotech_2007_01,Li_Opt_Comm_2011_01,Li_Opt._Expr_2012_01,Reddy_Nanotech_2016_01,Damilano_NL_2016_01} It is important to note that top-down methods are also suitable to obtain arrays of largely dislocation free NWs despite the presence of a high density of TDs in the layer used for the top-down process. In fact, the average number of TDs per NW is determined by the product of the TD density of the initial GaN layer and the cross-sectional area of one NW.\cite{Li_Opt_Comm_2011_01} For instance, for a substrate with a TD density on the order of 10$^{8}$~cm$^{-2}$, typical for GaN layers grown on Al$_{2}$O$_{3}$, and a NW diameter of 100~nm, less than 1\% of these objects will contain TDs.\cite{Wang2014}

The top-down fabrication of NW arrays typically relies on dry etching of a lithographically patterned two-dimensional layer. This fabrication method has, however, not achieved much popularity within the NW community because, in addition to the difficulties in achieving vertical NW sidewalls, the dry etching process inevitably creates point defects acting as nonradiative recombination centers at the NW sidewalls. Significant progress has been made in recent years by introducing an additional anisotropic wet chemical etching step after the creation of NWs by inductively coupled plasma reactive ion etching (ICP-RIE). This additional step removes the damaged material and facilitates the formation of smooth and straight NW sidewalls.\cite{Li_Opt_Comm_2011_01,Li_Opt._Expr_2012_01} Using this method, the fabrication of NW LEDs\cite{Li_Opt_Comm_2011_01} and optically pumped NW lasers has been demonstrated.\cite{Li_Opt._Expr_2012_01} As an alternative to this two-step etching approach, a new top-down fabrication method coined as selective area sublimation (SAS) was recently demonstrated by Damilano \emph{et al}.\cite{Damilano_NL_2016_01,Damilano_JCG_2017_01} This approach avoids any damage to the NW sidewalls as it is based not on  chemical etching but on material sublimation in vacuum, a process used before by different groups to decrease the diameter of as-grown NWs.\cite{Brockway_CGD_2011_01,Loitsch_Adv_Mat_2015_01,Zettler_NL_2016_01} Damilano \emph{et al} utilized this simple method to obtain random arrays of GaN NWs and (In,Ga)N/GaN NW heterostructures.\cite{Damilano_NL_2016_01,Damilano_JCG_2017_01}

For device applications, it is almost mandatory that the NWs do not have a random arrangement, but form an ordered array with precisely tunable spatial arrangements, NW diameters, and NW-to-NW spacings. In the present work, we explore the fabrication of ordered arrays of Ga-polar\cite{Zuniga-Perez_Appl_Phys_Rev_2016} GaN NWs by SAS. We demonstrate that this method is suitable to simultaneously fabricate ordered arrays of NWs with various diameters and spacings on a single wafer. The sublimation process, carried out under ultra high-vacuum conditions, is further analyzed \emph{in situ} by reflection high-energy electron diffraction (RHEED) and line-of-sight quadrupole mass spectrometry (QMS). A detailed analysis of the decomposition process allows us to assess the temperature dependence of the thermal etching rate, a crucial factor to control the formation by SAS of NWs as well as other types of nanostructures. Finally, we investigate the excitonic transitions from an ordered array of GaN NWs by low-temperature cathodoluminescence spectroscopy (CL). The emission from the array is found to be more intense than that from the original layer, confirming that the sublimation process does not introduce any nonradiative defects at the NW sidewalls, and revealing an enhanced extraction efficiency of the NW array. 

\section{Experimental}

\begin{figure*}
\centering
\includegraphics[width=0.9\textwidth]{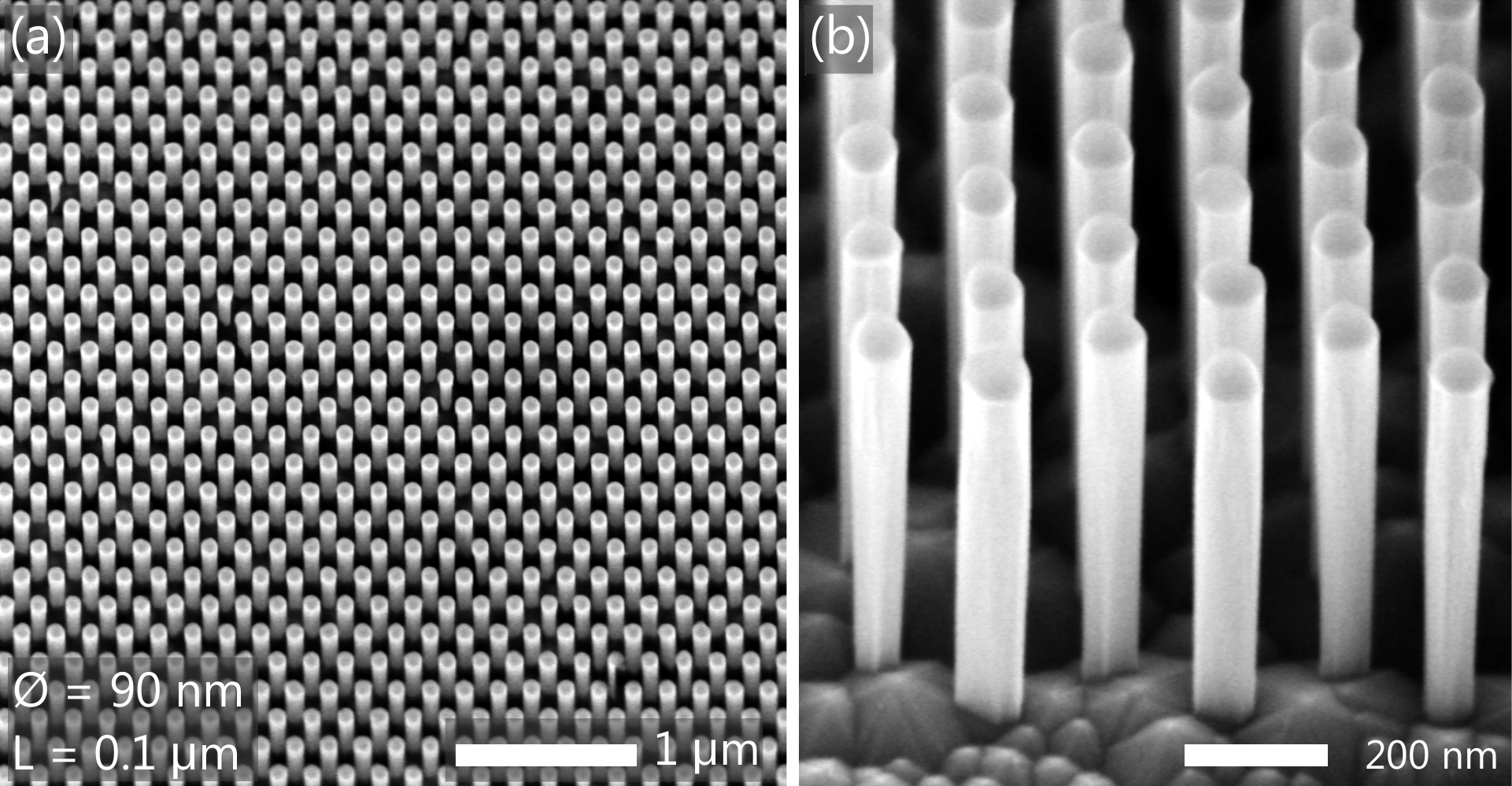}
\caption{\label{fig:SEM1} (a) and (b) Bird's eye-view scanning electron micrographs with different magnifications of a GaN NW array produced by selective area sublimation. The nominal values of the NW diameter ($\diameter$) and the spacing (L) are 90~nm and 0.1~µm, respectively. The micrograph shown in (b) is acquired at one of the edges of the patterned region.}
\end{figure*}

As substrates for SAS, we use commercially available $2^{\prime\prime}$  GaN(0001) layers grown by hydride vapor phase epitaxy on Al$_{2}$O$_{3}$ (purchased from Suzhou Nanowin Science and Technology). The thickness of the GaN layer is ($4.5\pm0.5$)~µm and its TD density $2\times10^{8}$~cm$^{-2}$. The backside of the substrates is coated with a 1.3~µm thick Ti layer for efficient thermal coupling during the sublimation process. To produce the mask for SAS, a 20~nm thick Si$_{x}$N$_{y}$ layer is first sputtered on the substrate surface. Afterwards, the Si$_{x}$N$_{y}$ layer is spin-coated with a 100~nm thick positive resist. Subsequently, the pattern is written by electron beam lithography (EBL) in a RAITH system. The pattern comprises $100\times100$~µm$^{2}$ fields containing hexagonal arrays of circles. Upon writing the pattern, the resist is developed before depositing 25~nm of Ni. After a lift-off step, the substrates are diced in $10\times10$~mm$^{2}$ pieces and the Si$_{x}$N$_{y}$ layer is etched at the regions not protected by Ni using RIE with a CF$_{4}$/O$_{2}$ gas mixture. Finally, the pieces are further etched by ICP-RIE using a mixture of Ar, Cl$_{2}$ and BCl$_{3}$ gases. This final etching step is used to eliminate residual surface contaminants created after etching Si$_{x}$N$_{y}$ as well as upon the chemical removal of the remaining Ni layer. 

The sublimation of the GaN layer is carried out under ultra high-vacuum conditions in a molecular beam epitaxy system where the process can be monitored \emph{in situ} by RHEED and QMS. The substrate temperature is measured with an emissivity corrected infrared optical pyrometer from Laytec. Customized Python routines are used to extract the angular intensity profiles around selected GaN diffraction spots of RHEED patterns. Because GaN decomposes congruently in vacuum \cite{Newman_JCG_1997}, we employ QMS to directly assess the decomposition rate by measuring the desorbing Ga flux. The QMS system response to the Ga$^{69}$ signal is calibrated in units of atoms/cm$^{2}$\,s using the procedure described in Ref.~\citenum{Fernandez-Garrido_NL_2015_01}. Regardless of the surface morphology, we assume that Ga atoms desorb isotropically.  

After the sublimation process, the samples are investigated by scanning electron microscopy (SEM) and low-temperature (14~K) CL spectroscopy and imaging. Scanning electron micrographs are acquired using either a Hitachi S-4800 or a Zeiss Ultra55 field-emission microscope. CL measurements are performed at acceleration voltages of 5~kV using a Gatan MonoCL4 system mounted to the Zeiss Ultra55 microscope. The system is equipped with a parabolic mirror for light collection and with both a photomultiplier and a charge-coupled device for detection.

\section{Results and discussion}
\subsection{Top-down fabrication of ordered arrays of GaN nanowires}

Ordered arrays of GaN NWs with different diameters and spacings in between NWs are produced simultaneously in an individual sample using a single sublimation step. Figure~\ref{fig:SEM1} shows scanning electron micrographs of an exemplary NW array produced by annealing a patterned substrate at 825\celsius for 20~min. The Si$_{x}$N$_{y}$ patches used as a mask for the sublimation process can be distinguished at the top of the GaN NWs by their darker contrast. The nominal NW diameter and spacing (distance between adjacent patches) values for this particular array are 90 and 100~nm, respectively. The final average NW diameter, which roughly corresponds to the actual size of the Si$_{x}$N$_{y}$ patches, is about 80~nm. As can be observed in Fig.~\ref{fig:SEM1}(a), the NW array is homogeneous on a large scale. Inside the patterned region, the NWs have rather vertical sidewalls [see Fig.~\ref{fig:SEM1}(b)]. The NW side facets are, however, not as flat and well defined as in the case of self-assembled GaN NWs produced by conventional bottom-up growth approaches.\cite{Trampert_Proc_2003_01,Largeau_nanotechnology_2008_01,Bergbauer_Nanotechnology_2010_01,Bertness_IEEE_2011_01,Geelhaar_IEEE_2011_01,Li_JAP_2012,Brandt_CGD_2014} Specifically, we do not observe pronounced \textit{M}-plane sidewall facets. The NW sidewalls are instead rather roundish, most likely because the Si$_{x}$N$_{y}$ patches do not have a hexagonal but a circular shape (a representative scanning electron micrograph of the Si$_{x}$N$_{y}$ patches before the thermal sublimation process is provided as supplemental material).

\begin{figure*}
\centering
\includegraphics[width=0.9\textwidth]{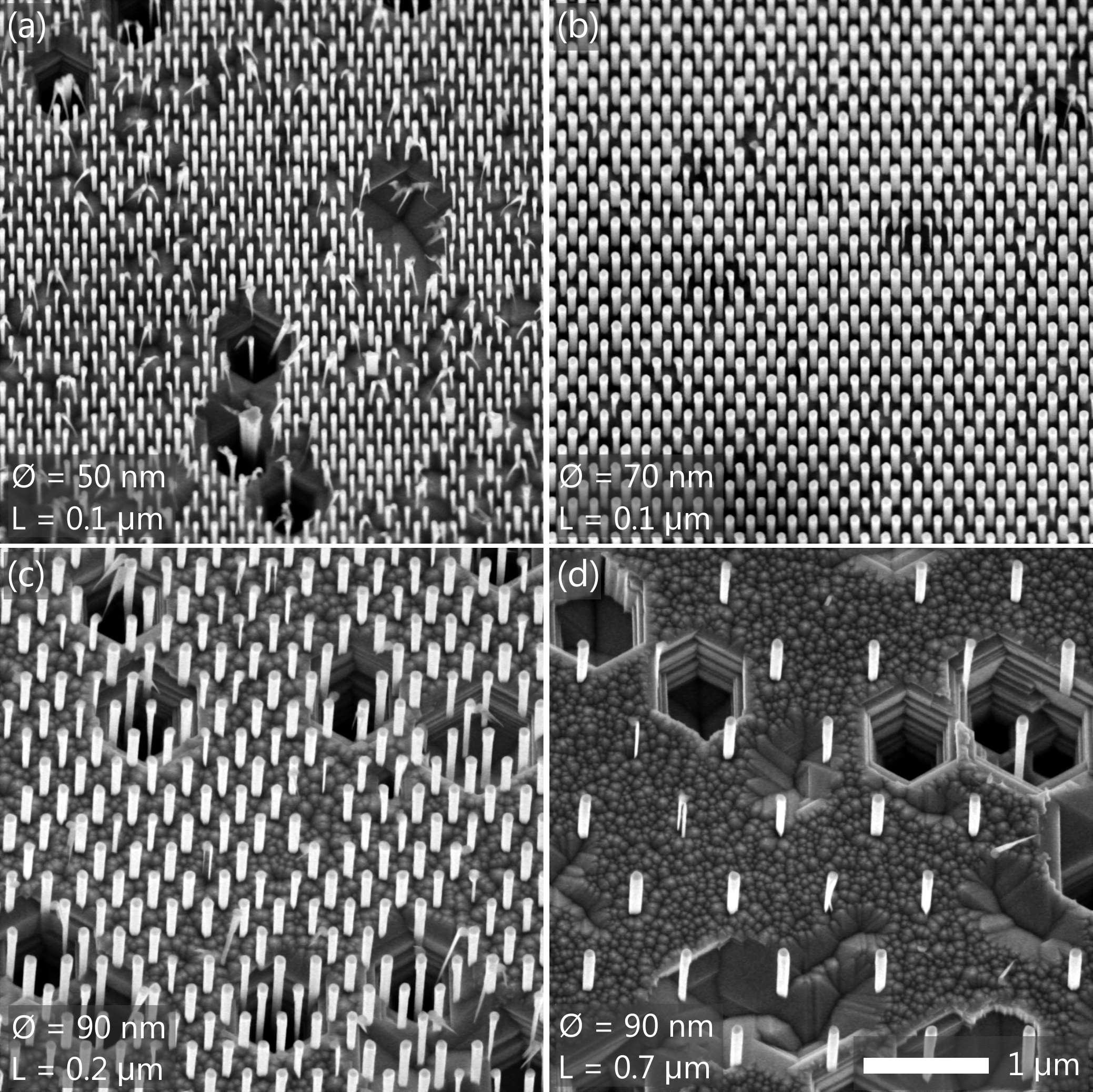}
\caption{\label{fig:SEM2} (a)--(d) Bird's eye view scanning electron micrographs of NW arrays produced by selective area sublimation in the same run with varying nominal NW diameter ($\diameter$) and spacing (L) values, as indicated in their corresponding labels. The magnification is the same for all micrographs and the scale bar is indicated in (d).}
\end{figure*}

Figure~\ref{fig:SEM2} presents scanning electron micrographs of four additional NW arrays produced at the same time as the one shown in Fig.~\ref{fig:SEM1}. These arrays differ from the previous one in either the NW diameter or the spacing. In all cases we observe the formation of homogeneous NW arrays with a very high yield. The yield only decreases due to the presence of TDs. The latter are easily recognized after the sublimation process because of the formation of hexagonal pits which are randomly distributed with a density of about $0.5 \times 10^{8}$~cm$^{-2}$, lower than the nominal TD density of the parent GaN layer $(2 \times 10^{8}$~cm$^{-2}$). This discrepancy originates likely from the fact that the sublimation process, analogously to chemical etching,\cite{Weyher_JCG_2007} preferentially etches screw TDs. Note that according to the nominal TD density of the GaN layer, the average number of TDs per NW (estimated as the product of the TD density of the original GaN layer and the NW cross-sectional area \cite{Li_Opt_Comm_2011_01}) amounts to 0.004 to 0.012 when the NW diameter is varied from 50 to 90~nm. Consequently, even for the array with the larger NW diameter, approximately 99\% of the NWs are expected to be free of TDs.\cite{Wang2014} The results shown in Figs.~\ref{fig:SEM1} and \ref{fig:SEM2} demonstrate that homogeneous arrays of GaN NWs with various diameters and spacings can be obtained with identical sublimation parameters.

\begin{figure*}
\centering
\includegraphics[width=0.9\textwidth]{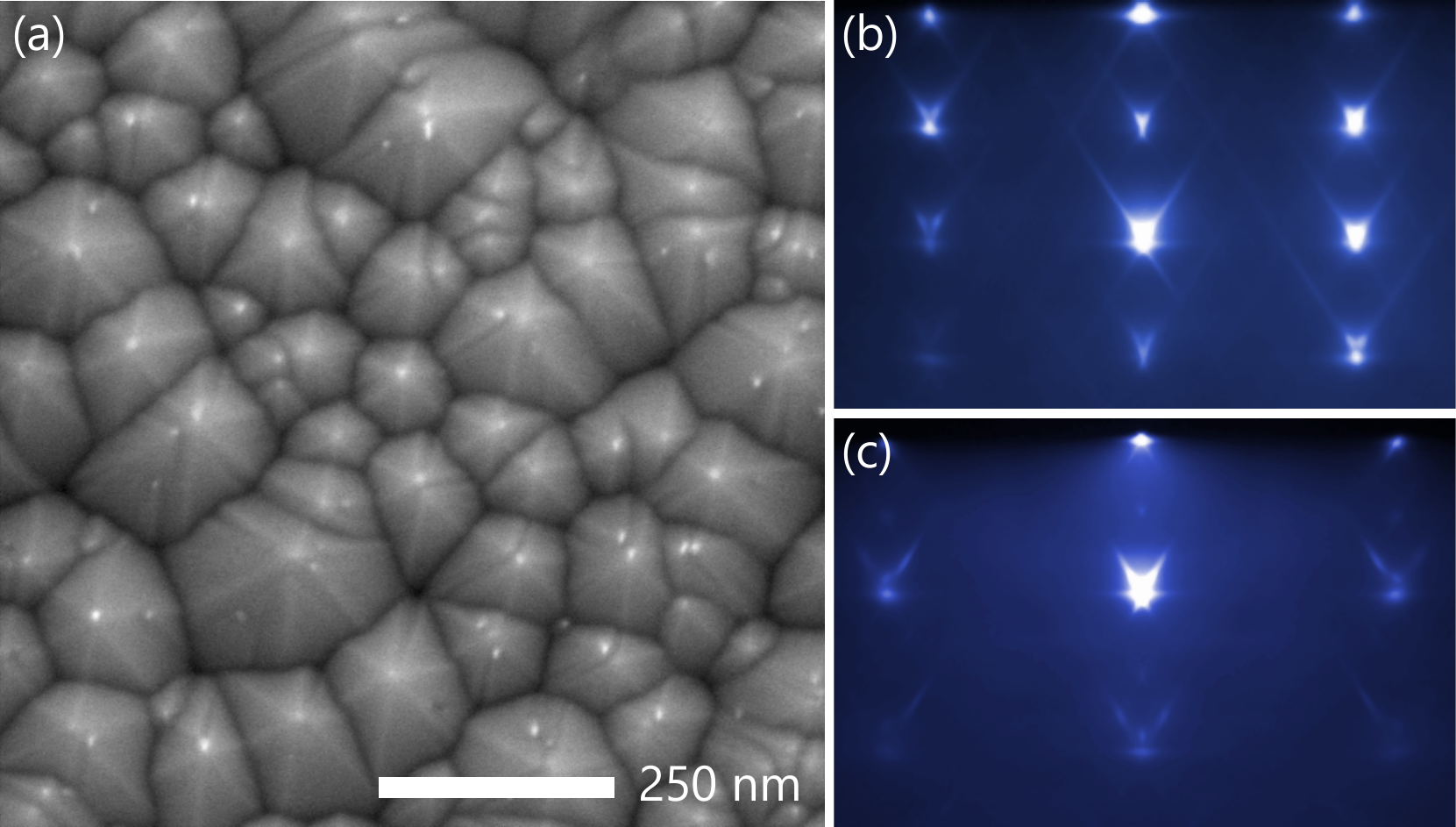}
\caption{\label{fig:SEM3} (a) Bird's eye view scanning electron micrograph of a GaN layer after 20~min of annealing in vacuum at 825\celsius. RHEED patterns acquired along the (b) $[11\bar{2}0]$ and (c) $[1\bar{1}00]$ azimuths after the thermal decomposition process.}
\end{figure*}

As can be seen in Figs.~\ref{fig:SEM2}(c) and \ref{fig:SEM2}(d), the morphology of the initially flat GaN layer in between the Si$_{x}$N$_{y}$ patches is clearly altered during the sublimation process. Figure~\ref{fig:SEM3}(a) presents a highly magnified scanning electron micrograph of an unpatterned GaN(0001) layer decomposed for 20~min at 825\celsius. This micrograph reveals that the original (0001) surface gives way to the formation of three-dimensional islands with a six-fold symmetry and well-defined semi-polar facets. This result is in apparent contrast to the step-edge and layer-by-layer decomposition mechanisms reported in Refs.~\citenum{Grandjean_APL_1999_01} and \citenum{Fernandez-Garrido_JAP_2008_04}, respectively. However, in these previous studies, where the decomposition process was analyzed at different temperatures, the GaN layer was not continuously decomposed (as in the present case), but a smooth (0001) surface was recovered prior to each (comparatively brief) decomposition step by depositing a thin GaN layer under conditions favoring step-flow growth. As a matter of fact, in Ref.~\citenum{Grandjean_APL_1999_01} it is also noted that continuous decomposition can result in surface faceting, a phenomenon that was tentatively ascribed to an enhanced sublimation rate near defects or/and grain boundaries. In the present study, the faceting under continuous decomposition is also detected \emph{in situ} by RHEED. Figures~\ref{fig:SEM3}(b) and \ref{fig:SEM3}(c) present the RHEED patterns recorded along the $[11\bar{2}0]$ and $[1\bar{1}00]$ azimuths, respectively. In both azimuths, a transmission pattern is observed, accompanied by pronounced chevrons. These features, caused by the refraction and transmission diffraction of electrons entering and exiting crystal facets, can be used to derive the shape of the three-dimensional objects from which they originate.\cite{Hanada_PRB_2001_01,Gutierrez_APL_2001_01,Pashley_Surf_Sci_2001_01,Gaire_Solid_Films_2009_01,Lee_Appl_Surc_Sc_2004_01,Ayahiko_Book_RHEED_2004} As discussed in detail in the supplemental material, the analysis of the vertex angles allows us to conclusively conclude that, in agreement with the results reported by Damilano \emph{et al} in Ref.~\citenum{Damilano_NL_2016_01}, the facets of the islands seen in Fig.~\ref{fig:SEM3} are formed by $\lbrace1\bar{1}03\rbrace$ planes.

\subsection{Thermal decomposition of GaN\textit{$\lbrace1\bar{1}03\rbrace$} facets: temperature dependence}
To properly control the fabrication of nanostructures by SAS, it is essential to know the precise temperature dependence of the GaN decomposition rate in vacuum. The decomposition of GaN(0001) layers has been measured as a function of the temperature by different groups.\cite{Groh_PSSA_1974_01,Held_SRL_1998_01,Ambacher_JVST_1996,Grandjean_APL_1999_01,Choi_SST_2002_01,Fernandez-Garrido_JAP_2008_04} However, as discussed in section 3.1, the (0001) surface is unstable against the formation of $\lbrace1\bar{1}03\rbrace$ facets, which are expected to decompose with a different rate. Hence, we next analyze the temperature dependence of the decomposition rate of $\lbrace1\bar{1}03\rbrace$ facets by measuring \emph{in situ} the desorbing Ga flux at different substrate temperatures. 

\begin{figure*}
\centering
\includegraphics[width=0.95\textwidth]{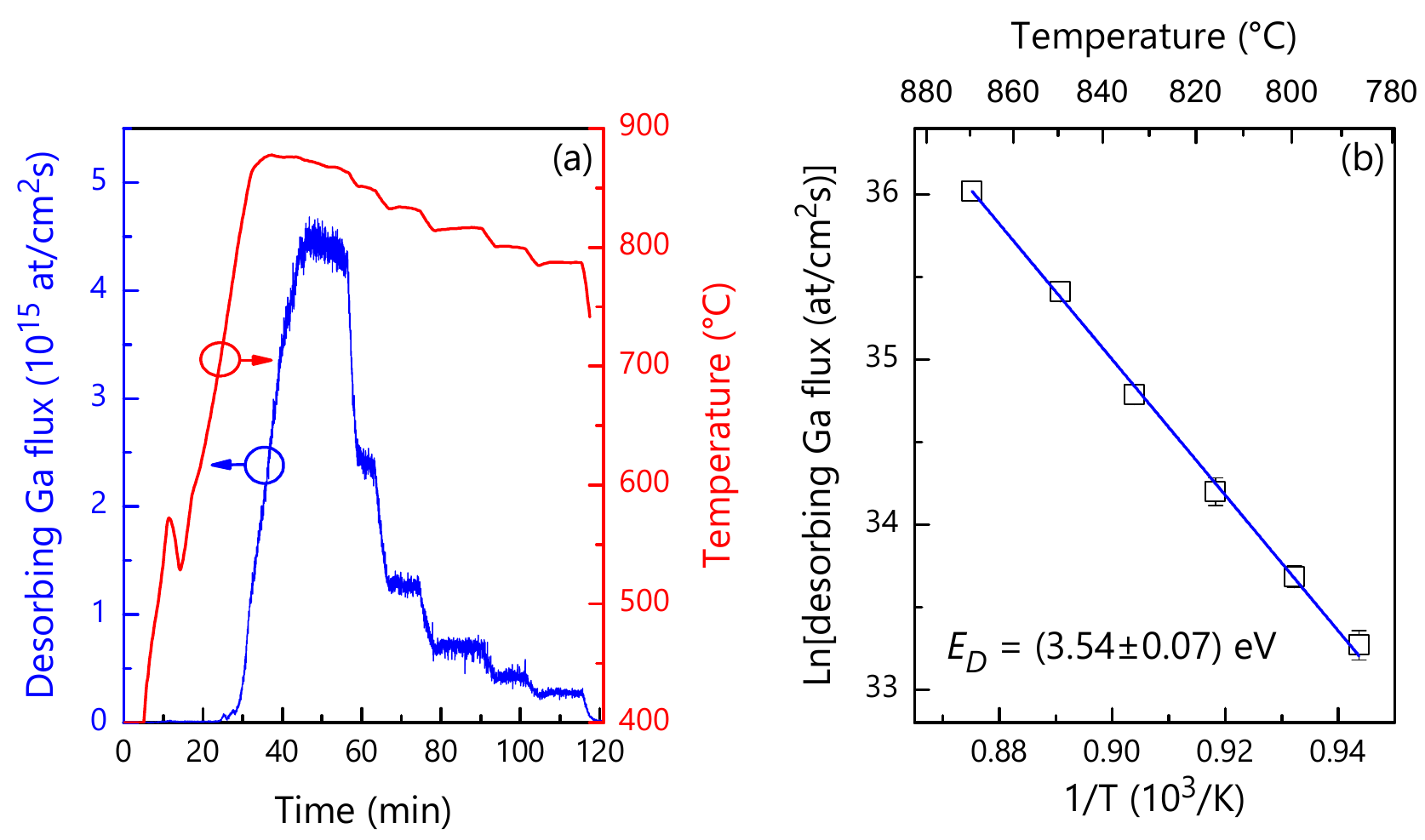}
\caption{\label{fig:QMS}(Color online) (a) Change of substrate temperature and the resulting temporal evolution of the desorbing Ga flux (red and blue lines, respectively) during the decomposition of a GaN layer. (b) Arrhenius representation of the temperature dependence of the desorbing Ga flux during the decomposition of a $\lbrace1\bar{1}03\rbrace$-faceted GaN layer. The line shows the best fit that yields an activation energy of $E_{D}=(3.54\pm0.07)$~eV.}
\end{figure*}

Figure~\ref{fig:QMS}(a) shows the desorbing Ga flux as measured by QMS during the congruent thermal decomposition of an unpatterned 2$^{\prime\prime}$ GaN wafer in vacuum. As shown in the graph, the substrate temperature $T$ is first increased up to 870\celsius with a rate of 20~K/min. Afterwards, the temperature is decreased in steps down to 786\celsius. To assess the desorbing Ga flux under-steady state conditions, we wait for the stabilization of the substrate temperature after every temperature step. The measurements evidence that the desorbing Ga flux steadily decreases from $4.4\times10^{15}$~cm$^{-2}$\,s$^{-1}$ ($\approx 60$~nm/min) at 870\celsius down to $0.3\times10^{15}$~cm$^{-2}$\,s$^{-1}$ ($\approx 4$~nm/min) at 786\celsius. The Arrhenius plot of the steady-steady state values of the decomposition rate of $\lbrace1\bar{1}03\rbrace$ facets ($\Phi_{D}$) is presented in Fig.~\ref{fig:QMS}(b). A fit of the data by
\begin{equation}
\label{Arrhenius}
\Phi_{D}=A \exp(-E_{D}/k_{B}T),
\end{equation}
with the Boltzmann constant $k_B$ yields a prefactor \textit{A} of $1.8 \times 10^{31}$~atoms cm$^{-2}$\,s$^{-1}$ and an activation energy \textit{E$_{D}$} of $(3.54 \pm 0.07)$~eV. When taking into account the increased surface area associated to the faceting, we find that the actual value of the exponential prefactor describing the thermal decomposition rate of $\lbrace1\bar{1}03\rbrace$ facets is reduced by a factor $(\tan(\theta)^{2}+1)^{1/2}$, where $\theta$ is the angle between the normal vectors of the $\lbrace1\bar{1}03\rbrace$ and $(0001)$ planes (32.0$^{\circ}$). The actual value of the exponential prefactor is thus $1.58 \times 10^{31}$~atoms cm$^{-2}$\,s$^{-1}$. The corresponding values of \textit{E$_{D}$} and \textit{A} measured by QMS in the case of a GaN(0001) plane are $(3.1\pm0.1)$~eV and $1.17~\times~10^{29}$~atoms cm$^{-2}$\,s$^{-1}$.\cite{Fernandez-Garrido_JAP_2008_04} Consequently, even though the energy barrier for thermal decomposition is slightly higher in the case of the $\lbrace1\bar{1}03\rbrace$ facets, in the temperature range of interest, these facets decompose faster than the $(0001)$ one due to the much higher exponential prefactor. These results are consistent with the idea that, while during growth the morphology of the crystal is governed by slow growing facets, during dissolution/thermal decomposition the crystal shape is dominated by fast desolving/decomposing crystal facets.\cite{Snyder_Materials_2007,Singh_CGD_2014} 

\subsection{Luminescence from ordered arrays of GaN nanowires produced by selective area sublimation}
\begin{figure*}
\centering
\includegraphics[width=\textwidth]{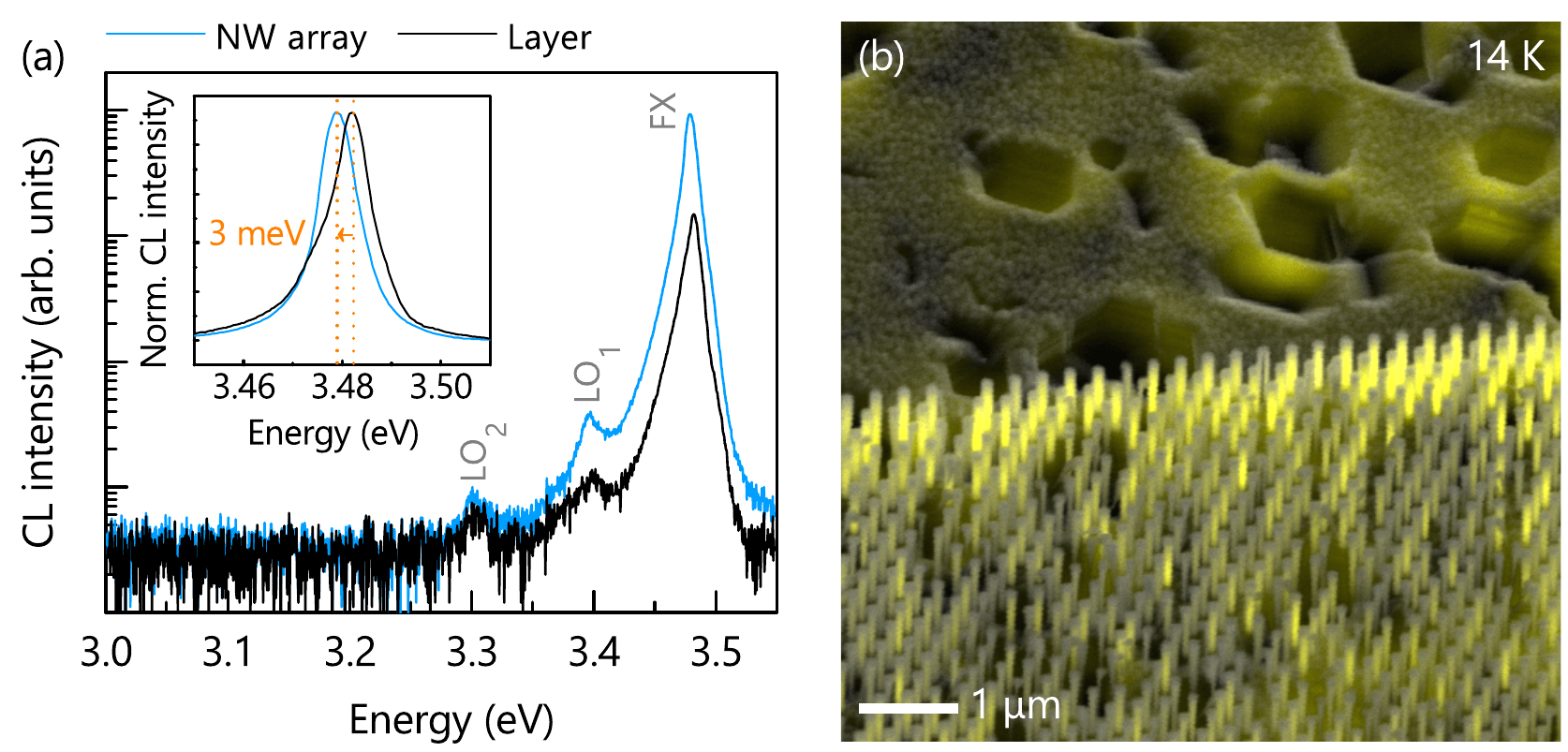}
\caption{\label{fig:CL}(Color online) (a) Near band-edge CL spectrum recorded at 14~K in bird's eye view geometry of the GaN NW array with a nominal NW diameter of 90~nm and a spacing of 0.1~µm. For comparison, the CL spectrum from the GaN layer surrounding the NW array (acquired under identical conditions) is also included. The emission from free excitons and its corresponding first and second order phonon replicas are labeled as FX, LO$_{1}$, and LO$_{2}$, respectively. The inset presents the normalized CL spectra of the NWs and the surrounding layer on a linear scale. (b) Superposition of a panchromatic CL intensity image recorded at 14~K (yellow) with the corresponding bird’s eye view scanning electron micrograph for the same NW array as in (a).}
\end{figure*}

The luminescence from ordered arrays of GaN NWs is investigated by CL spectroscopy at 14~K. Figure~\ref{fig:CL}(a) presents the near band-edge CL spectrum of the NW array shown in Fig.~\ref{fig:SEM1}. For comparison, we have also included a second CL spectrum recorded on an unpatterned region of the same sample. Both spectra are dominated by a high-energy line originating from the radiative decay of free excitons (FX), and its first- and second-order longitudinal-optical phonon replica (LO) at lower energies. The dominance of free over bound exciton recombination in this specific experiment is due to both the comparatively high excitation density (on the order of several $10^{17}$~cm$^{-3}$) and the resulting high effective carrier temperature [amounting to about 60~K for the spectra in Fig.\ref{fig:CL}(a), as determined by a fit to the high-energy slope of the free exciton line]. Two observations are worth to be stressed. First, the luminous intensity measured from the NW array is notably higher than that from the adjacent parent layer. This observation shows, most importantly, that the sublimation process does not degrade the internal quantum efficiency of the structure created, and furthermore, that the extraction efficiency of light is enhanced significantly over that of the planar reference.\cite{Reddy_Nanotech_2016_01,Hauswald_ACS_Phot_2017_01} Second, the spectrum of the NW array is rigidly red-shifted by 3--4~meV in comparison to the planar reference, as shown in the inset of Fig.~\ref{fig:CL}(a) for the FX transition. This redshift results from the elastic relaxation of the residual compressive strain in the parent GaN(0001) layer grown on Al$_{2}$O$_{3}$. The same phenomenon, benefiting from the large surface-to-volume ratio of these nanostructures, was observed by different groups for pure GaN NWs as well as in the case of compressively strained (In,Ga)N quantum wells embedded into GaN NWs.\cite{Chiu_Nanotech_2007_01,Hsieh_IEEE_EDL_2008_01,Ramesh_JAP_2010_01,Damilano_NL_2016_01,Damilano_JCG_2017_01}

The CL spectra shown in Fig.~\ref{fig:CL} were acquired immediately after exposing the sample to the electron beam. After prolonged exposure, the CL intensity in the patterned areas invariably quenches strongly. As an example, Fig.~\ref{fig:CL}(b) shows a panchromatic CL map superimposed with its corresponding bird's eye view scanning electron micrograph. The CL from the patterned region appears to be inhomogeneous with several NWs being apparently not emitting at all. However, this quenching of the CL intensiy upon irradation is solely the result of carbonaceous depositions that introduce nonradiative recombination channels at the NW sidewalls, as discussed in detail in Ref.~\citenum{Laehnemann_Nanotech_2016_01}. 

\section{Summary and conclusions}
We have demonstrated the fabrication of ordered arrays of GaN NWs by SAS of pre-patterned GaN(0001) layers grown on Al$_{2}$O$_{3}$. In a single sample, we simultaneously produced different arrays with NW diameters and spacings ranging from 50 to 90~nm and 0.1 to 0.7~µm, respectively. The resulting NW sidewalls are vertical, but do not exhibit well-defined \textit{M}-plane facets. The roundish shape of the NWs is attributed to the use of Si$_{x}$N$_{y}$ patches with a circular shape as a mask for SAS. According to Ref.~\citenum{Damilano_NL_2016_01}, we expect the formation of \textit{M}-plane sidewall facets when employing patches with a hexagonal shape properly oriented with respect to the GaN template underneath. During the sublimation process, the (0001) surface vanishes giving way to the formation of fast decomposing semi-polar $\lbrace1\bar{1}03\rbrace$ facets. The stability of these facets determines the thermal etching rate. We found that the $\lbrace1\bar{1}03\rbrace$ facets decompose following an Arrhenius-like temperature dependence with an activation energy of $(3.54\pm0.07)$~eV and an exponential prefactor of $9.46\times10^{32}$~atoms\,cm$^{-2}$\,s$^{-1}$. Low-temperature CL experiments reveal a higher luminous intensity from the NW array thanks to an improved light extraction efficiency. The emission is red-shifted with respect to the one of the GaN layer because the large NW aspect ratio facilitates the elastic relaxation of residual strain. Although our CL experiments indicate that the sublimation process does not generate nonradiative recombination centers at the NW sidewalls, more conclusive results in this respect could be obtained by time-resolved photoluminescence spectroscopy. Such an analysis would be possible for GaN NWs containing (In,Ga)N quantum wells, for which the luminescence signals from the NWs and the GaN layer underneath would not spectrally overlap.

Selective area sublimation is, therefore, a suitable top-down approach to produce ordered arrays of GaN NWs with high luminous efficiency without the need of any elaborate chemical treatment. This fabrication method could be readily extended to other types of micro- and nanostructures as well as to additional material systems provided that they decompose congruently (to avoid the accumulation of constituent elements on the sample surface) and exhibit a marked anisotropy in the stability of its crystallographic planes. A prominent candidate for such experiments is ZnO, another wide-bandgap semiconductor of interest for optical applications that shares many properties with GaN. 

\section*{Conflicts of interest}
There are no conflicts to declare.

\section*{Acknowledgements}
We thank Katrin Morgenroth for her support during the preparation and characterization of the samples as well as for her dedicated maintenance of the molecular beam epitaxy system together with Carsten Stemmler and Hans-Peter Schönherr, Sebastian Meister and Sander Rauwerdink for patterning the substrates, and Anne-Kathrin Bluhm for her help during the acquisition of scanning electron micrographs. We are indebted to David van Treeck for numerous discussions on SAS, and to Vladimir Kaganer for discussions on the probability to find dislocations in NWs. Special thanks are due to Lutz Geelhaar for his continuous encouragement and support and a critical reading of the manuscript. Funding from the Bundesministerium für Bildung und Forschung through project FKZ:13N13662 is gratefully acknowledged. Sergio Fernández-Garrido acknowledges the partial financial support received through the Spanish program Ramón y Cajal (co-financed by the European Social Fund) under grant RYC-2016-19509 from Ministerio de Ciencia, Innovación y Universidades.



\balance


\bibliography{bibliography} 
\bibliographystyle{rsc} 

\clearpage

\renewcommand{\figurename}{Table of contents entry}
\setcounter{figure}{0}
\renewcommand{\thefigure}{}%

\section*{Table of contents entry}
\begin{figure}
\centering
\includegraphics[width=7.7cm]{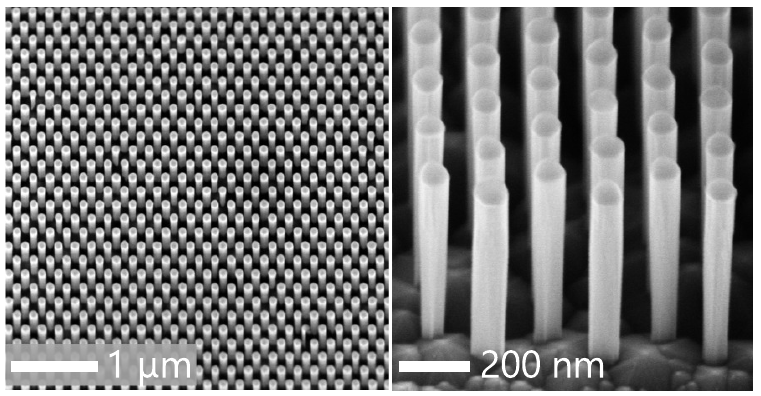}
\caption{\label{fig:TOC} We demonstrate the top-down fabrication of ordered arrays of GaN nanowires by selective area sublimation of pre-patterned GaN(0001) layers.}
\end{figure}

\end{document}